\documentclass{DISproc}

\newcommand{\Tr}{\mbox{Tr}}

\begin{document}
\title{Subleading-$N_c$ improved parton showers
\hfill\begin{small}DESY 12-075\end{small}}

\author{{\slshape Simon Pl\"atzer$^1$, Malin Sj\"odahl$^2$}\\[1ex]
$^1$DESY, Notkestra{\ss}e 85, 22607 Hamburg, Germany\\
$^2$Lund University,
  S\"olvegatan 14A, SE-223 62 Lund, Sweden }

\contribID{xy}

\doi  

\maketitle

\begin{abstract}
We present an algorithm for improving subsequent parton shower
emissions by full SU(3) colour correlations in the framework of a
dipole-type shower.  As a proof of concept, we present results from
the first implementation of such an algorithm for a final state
shower.
\end{abstract}


\section{Introduction and motivation}

Parton showers and event generators are indispensable tools for
predicting and understanding collider results
\cite{Sjostrand:2007gs,Bahr:2008pv,Gleisberg:2003xi}.  Considering
their importance for interpreting LHC results, it is essential to have
a good understanding of their approximations and limitations.  These
simulations have up to now all been based on QCD as an $SU(N_c)$ gauge
theory in the limit of large $N_c$. For $N_c=3$, this approximation
seems to work remarkably well despite the fact that $1/N_c=1/3$, in
the general case, and $1/N_c^2=1/9$ in most cases are not truly small
parameters. Otherwise, significant deviations from parton shower
predictions as compared to experimentally measured observables would
have already hinted towards a severe underestimate of colour
suppressed terms.

Including colour suppressed terms in parton shower simulations has so
far been an unexplored continent, and investigating effects caused by
colour correlations beyond the large-$N_c$ limit is mandatory in the
age of ever improving simulations, particularly when including
higher-order QCD corrections. Especially when considering the matching
of parton showers to NLO QCD corrections, subleading-$N_c$ improved
parton showers provide a valuable input in making these matchings more
precise such that the matching conditions are indeed satisfied
exactly, and not only modulo colour suppressed terms.\footnote{In this
  context an independent approach, considering only one emission, has
  been presented in \cite{Hoeche:2011fd}.}

We here present an approach to subleading colour contributions
\cite{Platzer:2012qg} which is simple in the sense that it fits very well into
the framework of existing Monte Carlo event generators. We note that
this is not the end of the story, as for the general case an
evolution at amplitude level would have to be considered. Our approach
of colour matrix element corrections is a first step towards
quantifying the size of the expected effects.

\section{From dipole factorization to dipole showering}

Dipole factorization, 
\cite{Catani:1996vz},
states that the behaviour of QCD tree-level matrix elements squared in
any singly unresolved limit involving two partons $i,j$ ({\it i.e.}
whenever $i$ and $j$ become collinear or one of them soft), can be
cast into the form
\begin{multline}
  \label{eqs:dipolefactorization}
  |{\cal M}_{n+1}(...,p_i,...,p_j,...,p_k,...)|^2 \approx\\
  \sum_{k\ne i,j} \frac{1}{2 p_i\cdot p_j}
  \langle {\cal M}_n(...,p_{\tilde{ij}},...,p_{\tilde{k}},...) |
          {\mathbf V}_{ij,k}(p_i,p_j,p_k)| {\cal M}_n(...,p_{\tilde{ij}},...,p_{\tilde{k}},...)\rangle \ ,
\end{multline}
where $|{\cal M}_{n}\rangle$ -- which is a vector in the space of
helicity and colour configurations -- denotes the amplitude for an
$n$-parton final state. Here an emitter $\tilde{ij}$ undergoes
splitting to two partons $i$ and $j$ in the presence of a spectator
$\tilde{k}$ which absorbs the longitudinal recoil of the splitting,
$\tilde{k}\to k$. This factorization formula, which is well
established to provide a subtraction scheme for NLO calculations, can
actually be used to derive a dipole-type shower algorithm
\cite{Platzer:2009jq}. Results of an implementation have been reported
in \cite{Platzer:2011bc}, and similar approaches have been considered in
\cite{Dinsdale:2007mf,Schumann:2007mg}.
In these cases, the colour correlations present in ${\mathbf
  V}_{ij,k}$ are approximated in the large-$N_c$ limit, while keeping
the colour factor for gluon emission off quarks, $C_F={\mathbf
  T}_{q_i}^2$, exact. In turn, chains of colour connected dipoles
evolve through subsequent emissions generating more dipoles in a chain
or leading to a breakup of the chain in case a gluon splits into a
$q\bar{q}$ pair until eventually the transverse momentum of potential
emissions is below an infrared cutoff in the region of one GeV.

\section{Colour matrix element corrections}

The dipole factorization formula eq.~\ref{eqs:dipolefactorization}
implies a factorization at the level of cross sections; here, the
differential cross section for $n+1$ partons factorizes into the cross
section for producing $n$ partons times a radiation density as the sum
over all dipole configurations $\tilde{ij},\tilde{k}$, which undergo
radiation, $\tilde{ij},\tilde{k}\to i,j,k$:
\begin{equation}
  \label{eqs:cmec}
        dP_{ij,k}(p_\perp^2,z) =
        V_{ij,k}(p_\perp^2,z)\frac{d\phi_{n+1}(p_\perp^2,z)}{d\phi_{n}}
         \times
        \frac{-1}{{\mathbf T}_{\tilde{ij}}^2} \frac{\langle
          {\cal M}_n|{\mathbf T}_{\tilde{ij}}\cdot {\mathbf T}_k |{\cal M}_n\rangle
        }{|{\cal M}_n|^2}
\end{equation}
Here, we have used the spin-averaged version of the dipole kernels
including the product of colour charges encoding the colour
correlations, ${\mathbf V}_{ij,k}= -V_{ij,k} {\mathbf
  T}_{\tilde{ij}}\cdot {\mathbf T}_k/{\mathbf T}_{\tilde{ij}}^2$, and
${\rm d}\phi_{k}$ denotes the $k$-parton phase space. In the
large-$N_c$ limit, this formula yields the basis for the dipole shower
considered so far: $-{\mathbf T}_{\tilde{ij}}\cdot {\mathbf
  T}_k/{\mathbf T}_{\tilde{ij}}^2 \to \delta(\tilde{ij}, k \text{
  colour connected})/(1+\delta_{\tilde{ij}})$, where
$\delta_{\tilde{ij}}=1(0)$ for $\tilde{ij}=g(q/\bar{q})$. To obtain an
algorithm which instead keeps the full colour correlations, we do not
consider this approximation but keep the second factor in
eq.~\ref{eqs:cmec} exactly. Owing to the similarity of matrix element
corrections present in parton showers so far, we refer to this
improvement as `colour matrix element corrections'.

Eq.~\ref{eqs:cmec} describes how a single emission incorporates colour
correlations. Indeed, for the first emission off the hard subprocess,
$|{\cal M}_n\rangle$ is known, though it has to be recalculated after
each subsequent emission to define the colour matrix element
correction for the next emission. Instead of directly calculating the
next amplitude, which would only be possible if we had derived
splitting amplitudes in the singly unresolved limits, we observe that
\begin{equation}
  |{\cal M}_n|^2 = {\cal M}_n^\dagger S_n {\cal M}_n =
  \Tr \left( S_n\times {\cal M}_n{\cal M}_n^\dagger \right) 
\end{equation}
and
\begin{equation}
  \langle {\cal M}_n|{\mathbf T}_{\tilde{ij}}\cdot {\mathbf T}_{\tilde{k}}|{\cal M}_n\rangle = 
  \Tr \left( S_{n+1}\times T_{\tilde{k},n} {\cal M}_n{\cal M}_n^\dagger T_{\tilde{ij},n}^\dagger \right) \ ,
\end{equation}
where we have chosen a definite basis $\{|\alpha\rangle\}$ for the colour space,
$
  |{\cal M}_n\rangle = \sum_{\alpha=1}^{d_n} c_{n,\alpha} |\alpha_n\rangle
  \quad \leftrightarrow \quad {\cal M}_n = (c_{n,1},...,c_{n,d_n})^T
$
and introduced the scalar product matrix
$S_n=\{\langle\alpha_n|\beta_n\rangle\}$ as well as matrix
representations of the colour charge operators for $n$ partons, ${\mathbf T}_i\to T_{i,n}$.

These representations then imply that we can work with an amplitude
matrix $M_n$ as the fundamental object,
\begin{equation}
  M_{n+1} =
  -\sum_{i\ne j}\sum_{k\ne i,j} \frac{4\pi\alpha_s}{p_i\cdot p_j} \frac{V_{ij,k}(p_i,p_j,p_k)}{{\mathbf T}_{\tilde{ij}}^2}
  \ T_{\tilde{k},n}M_n T_{\tilde{ij},n}^\dagger \ ,
\end{equation}
where the initial matrix for the hard subprocess is given by $
  M_{n_\text{hard}} = {\cal M}_{n_\text{hard}}{\cal M}_{n_\text{hard}}^\dagger
$.

\section{Technicalities}

Having outlined the principle of the algorithm for including colour
correlations for subsequent parton shower emissions, two major
technical issues have to be addressed: On the one hand, an general
treatment of the colour basis for an arbitrary number of partons is
required. On the other hand, sampling from the probability
(Sudakov-type) density driving the next parton shower emission has to
be generalized to the case of non-positive splitting rates as
typically encountered for $1/N_c$ suppressed contributions.

For the first task, we have implemented a C++ library
\textsf{ColorFull} \cite{Sjodahl:Colorfull} implementing the trace
bases of colour space, \cite{Sjodahl:2009wx}. This library is
interfaced to the \textsf{Matchbox} framework presented in
\cite{Platzer:2011bc}. The colour matrix element corrections
calculated in this part of the simulations, are inserted as correction
weights into an existing dipole shower implementation, which uses the
\textsf{ExSample} library \cite{Platzer:2011dr} to sample Sudakov-type
densities derived from the absolute value of the colour-corrected
splitting rates. For the second task, we then employ the interleaved
competition/veto algorithm outlined in \cite{Platzer:2011dq} to arrive
at events distributed according to the desired density (note that the
sum of all splitting rates approximates a squared matrix element and
is thus positive).

\section{Results}

As a proof-of-concept, we present results from the subleading-$N_c$
improved parton shower for final state radiation, more precisely
considering $e^+e^-\to q\bar{q}$ at LEP1 energies. We compare three
different approximations: `full' colour correlations, the `shower'
approximation where the ${\mathbf T}_i\cdot {\mathbf T}_j/{\mathbf
  T}_i^2$ are taken in the large-$N_c$ limit, and a `strict'
large-$N_C$ approximation, where also $C_F\approx
C_A/2$. Interestingly, the `shower' approximation does not exactly
reproduce the shower implementation; from four partons onwards, this
evolution is sensitive to the emission history, though matches the
shower implementation if one sequence of dipoles has
dominated. Indeed, the differences between the `shower' approximation
and the default shower implementation are at the per-mille level as
one would expect from strong ordering in the emission history.

\begin{figure}
\begin{center}
\includegraphics[scale=0.6]{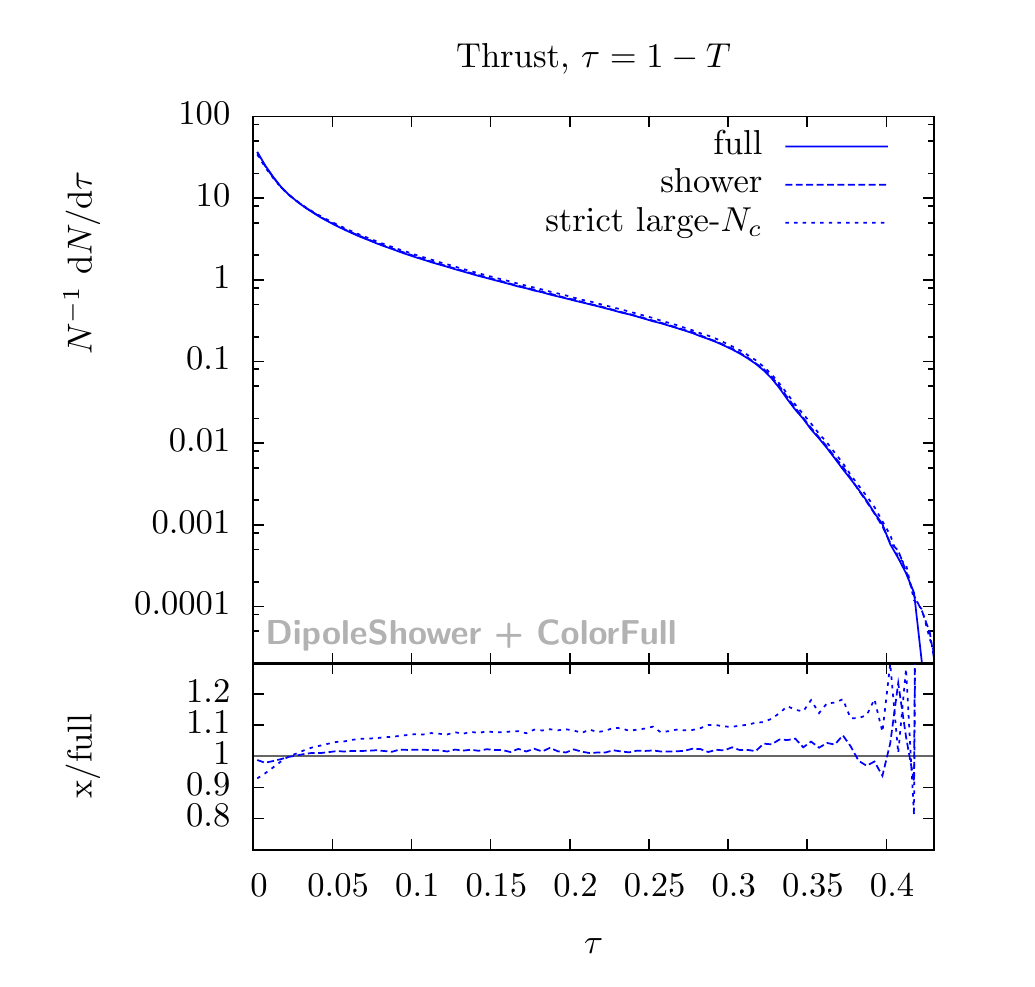}
\includegraphics[scale=0.6]{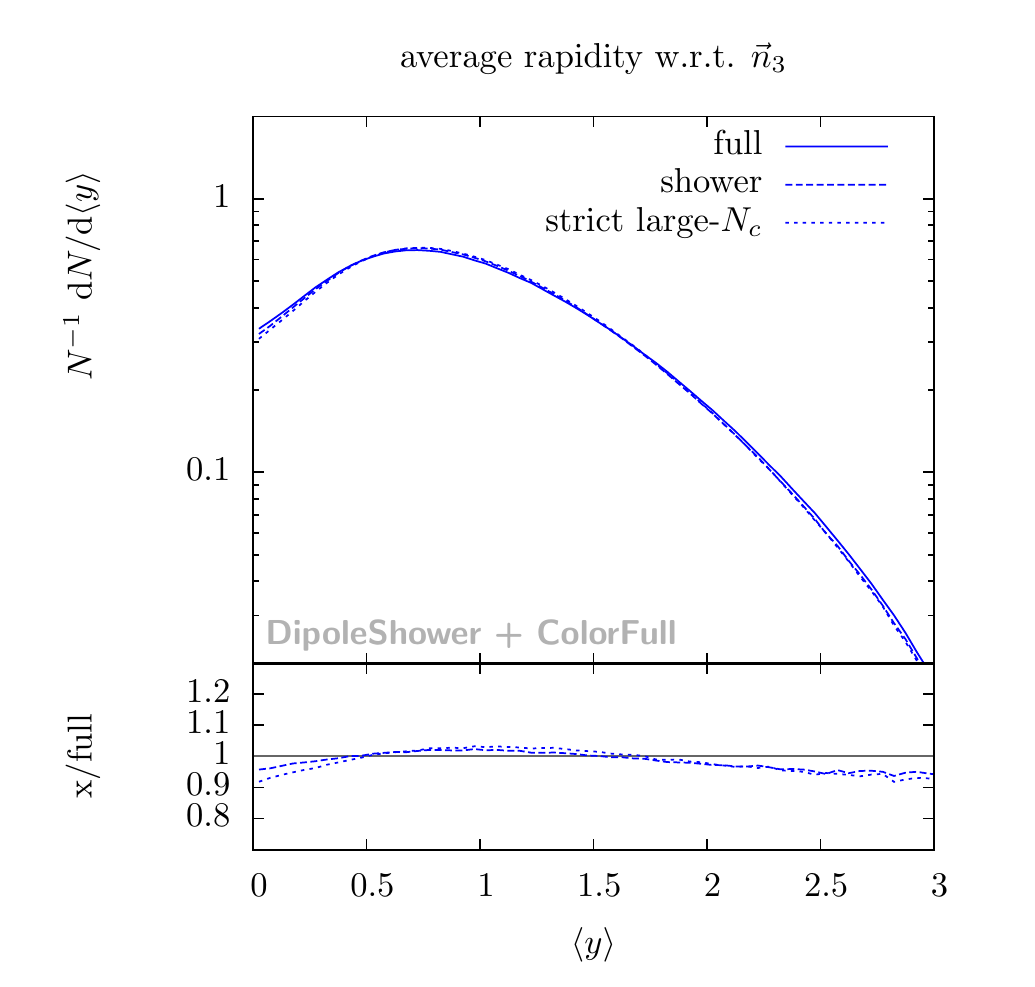}
\end{center}
\caption{\label{figs:sampleresults}The thrust distribution (left) and
  the average rapidity w.r.t the thrust axis defined by the three hardest
  partons using the different approximations.}
\end{figure}

In the results presented here, we include up to six improved shower
emissions. $g\to q\bar{q}$ splittings are neglected, as there are no
associated colour correlations; we also do not include hadronization.
For event shapes and jet rates we find small subleading-$N_c$ effects
when considering the shower approximation; for tailored observables,
probing the dynamics of soft radiation with respect to a hard
subsystem of the event, larger effects are seen. The strict
approximation shows larger deviations. This fact can mainly be
attributed to the change in the Sudakov exponent for gluon emission
off a quark.  A few sample results are shown in
figure~\ref{figs:sampleresults}.

\section{Conclusions}

We have presented the first implementation of a subleading-$N_c$
improved parton showers. For $e^+e^-\to \text{jets}$ small effects are
seen except for very special observables. The technical issues
associated with the implementation, particularly the treatment of the
colour basis and the presence of negative splitting kernels will serve
as input for related and future work; we also anticipate that larger
effects can be seen in hadron collisions, {\it e.g.} $pp\to
\text{jets}$.

\begin{footnotesize}
\bibliographystyle{DISproc}
\bibliography{platzer_simon}
\end{footnotesize}

\end{document}